# Deducing radiation pressure on a submerged mirror from the Doppler shift

Masud Mansuripur

College of Optical Sciences, The University of Arizona, Tucson, Arizona 85721

<masud@optics.arizona.edu>

**Abstract**. Radiation pressure on a flat mirror submerged in a transparent liquid, depends not only on the refractive index $n$ of the liquid, but also on the phase angle $\psi_o$ of the Fresnel reflection coefficient of the mirror, which could be anywhere between 0° and 180°. Depending on the value of $\psi_o$, the momentum per incident photon picked up by the mirror covers the range between the Abraham and Minkowski values, i.e., the interval $(2\hbar\omega_o/nc, 2n\hbar\omega_o/c)$. Here $\hbar$ is the reduced Planck constant, $\omega_o$ is the frequency of the incident photon, and $c$ is the speed of light in vacuum. We argue that a simple experimental setup involving a dielectric slab of refractive index $n$, a vibrating mirror placed a short distance behind the slab, a collimated, monochromatic light beam illuminating the mirror through the slab, and an interferometer to measure the phase of the reflected beam, is all that is needed to deduce the precise magnitude of the radiation pressure on a submerged mirror. In the proposed experiment, the transparent slab plays the role of the submerging liquid (even though it remains detached from the mirror at all times), and the adjustable gap between the mirror and the slab simulates the variable phase-angle $\psi_o$. The phase of the reflected beam, measured as a function of time during one oscillation period of the mirror, then provides the information needed to determine the gap-dependence of the reflected beam's Doppler shift and, consequently, the radiation pressure experienced by the mirror.

**1. Introduction**. The 1954 experiments of Jones and Richards on submerged mirrors [1], followed by those of Jones and Leslie nearly a quarter century later [2,3], established the dependence of radiation pressure on the refractive index $n$ of the submerging liquid. The radiation pressure found in these experiments was proportional to $n$, thus providing strong evidence in favor of the Minkowski momentum of the photon inside dielectric media. Later calculations of the Lorentz force exerted on submerged mirrors showed the pressure to depend not only on the refractive index $n$ of the liquid, but also on the phase-angle $\psi_o$ of the Fresnel reflection coefficient of the mirror [4,5]. Specifically, whereas the pressure on a mirror having $\psi_o = 180°$ is directly proportional to $n$, that on a mirror with $\psi_o = 0°$ is inversely proportional to $n$. A continuous variation of $\psi_o$ between these extreme values would result in a continuous change of the pressure experienced by the mirror, indicating, at first sight, a photon momentum that could be anywhere between the Abraham value of $\hbar\omega_o/nc$ and the Minkowski value of $n\hbar\omega_o/c$. Here $\hbar$ is the reduced Planck constant, $\omega_o$ is the frequency of the incident photon, and $c$ is the speed of light in vacuum. Since most conventional mirrors have a Fresnel reflection coefficient close to $-1.0$, it is reasonable to assume that the phase of the mirrors used in the aforementioned experiments must have been close to 180°, which explains the observation of the Minkowski momentum in those experiments.

An alternative explanation of the submerged mirror experiments that emerges from Lorentz force calculations [6-15] is as follows. Inside the liquid, in the region between the light source and the mirror, the beam reflected from the mirror interferes with the incident beam. The Lorentz force that the resulting interference fringes exert on the liquid tends to pull the liquid away from the mirror when $\psi_o = 180°$, and it tends to push the liquid toward the mirror when $\psi_o = 0°$. The total force (acting on both the mirror and the liquid) is always the same, regardless of the value of $\psi_o$, and the photon momentum in the liquid is neither Abraham's nor Minkowski's, but is given by $\tfrac{1}{2}(n+n^{-1})\hbar\omega_o/c$, which is part electromagnetic ($\hbar\omega_o/nc$) and part mechanical [$\tfrac{1}{2}(n-n^{-1})\hbar\omega_o/c$] in nature [5,11].

The present paper provides a different kind of evidence in support of the above picture painted by the Lorentz force calculations of radiation pressure on submerged mirrors. The basis of the new argument, Doppler shift of the light reflected from a moving mirror, is fundamentally different from that of the Lorentz force, yet the conclusions reached by the two arguments are identical. We describe in Sec.2 the proposed system for monitoring the Doppler shift of a light beam upon reflection from a vibrating mirror placed immediately behind a transparent dielectric slab. The expected value of the Doppler shift (based on theoretical considerations) is calculated in Sec.3, and related to the radiation pressure on the mirror in Sec.4. In Sec.5, we directly calculate the radiation pressure on the mirror behind the slab, and show that the result coincides with that obtained from the Doppler shift argument. Some final thoughts and concluding remarks are the subject of Sec.6.

**2. Illuminating a flat mirror through a dielectric slab**. With reference to Fig.1, consider a pulse of light having energy $\mathcal{E}$, duration $\tau$, and center frequency $\omega_o$, arriving at normal incidence on a flat mirror through a thick dielectric slab of refractive index $n$. The front facet of the slab is anti-reflection coated to eliminate reflection losses, but also, more significantly, to prevent multiple reflections within the slab. The mirror has a Fresnel reflection coefficient $\rho\exp(i\psi)$, with $\rho\approx 1$ and $\psi$ being a fixed but arbitrary parameter somewhere between 0° and 180°. We also allow for a small air-gap $d$ between the mirror and the rear facet of the dielectric slab.

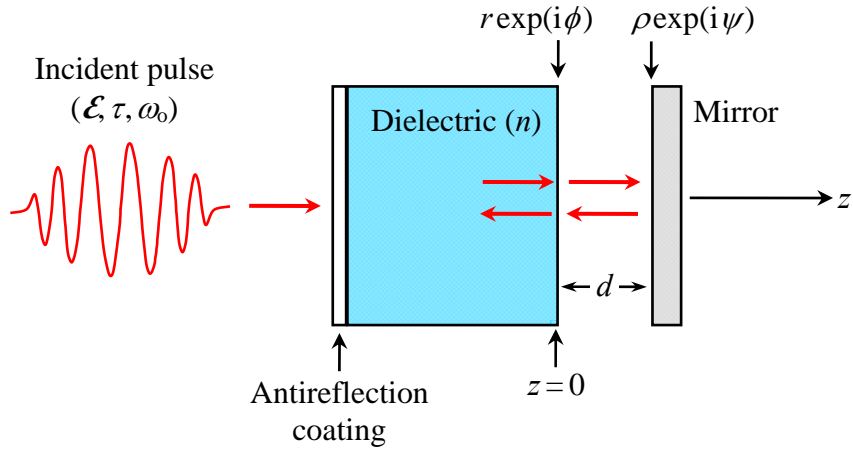

**Fig.1** (Color online). A mirror having Fresnel reflection coefficient $\rho\exp(i\psi)$ is placed a short distance $d$ behind a fairly thick dielectric slab of refractive index $n$. The incident light pulse propagating along the $z$-axis has energy $\mathcal{E}$, duration $\tau$, frequency $\omega_o$, and wavelength $\lambda_o = 2\pi c/\omega_o$. The Fresnel reflection coefficient of the dielectric surface at $z=0$, denoted by $r\exp(i\phi)$, is a function of $n$, $\rho$, $\psi$, $d$ and $\omega_o$.

The air-gap need not be large; in fact the range of values of $d$ that is of interest here is $0 \leq d \leq \lambda_o/4$, where $\lambda_o = 2\pi c/\omega_o$ is the vacuum wavelength of the incident light; $c$ is the speed of light in vacuum. The gap has practical value in that it allows one to measure the pressure on the mirror without having to worry about the dielectric slab sticking to the mirror. This eliminates the possibility of measurement errors due to the viscous force as well as the force of radiation that acts on the slab. Another advantage of the air-gap is that, by adjusting $d$, one can shift the effective phase of the Fresnel reflection coefficient of the mirror to any desired value. It will be seen in the final equations that $\psi$ and $d$ appear only in the combination $\psi + 4\pi d/\lambda_o$; therefore, the effective phase of the mirror used in any experiment can be adjusted between 0° and 180° simply by changing the air-gap.

Ideally, the light pulse should be short enough that any overlap between the incident and reflected pulses would occur within the dielectric slab (and, of course, in the air-gap as well), but not in the region preceding the slab. For a 10 ps pulse, for instance, a 1.0 cm-thick slab would be appropriate. This will simplify the computation of the Lorentz force on the slab, which would otherwise have to be extended into the antireflection coating layer(s). In principle, of course, there is no reason to limit the experiment to such short pulses; even a continuous-wave (CW) laser beam could be acceptable, provided that the temporal coherence length of the light source is short enough to confine the interference fringes – arising from overlap between incident and reflected beams – to the vicinity of the rear facet of the slab located at $z = 0$ [11]. The constraint should thus be properly placed not on the length of the pulse in the time domain, but on its spectral width in the Fourier domain. None of these considerations apply, however, if the goal of the experiment is to measure only the radiation pressure on the mirror, not that on the slab.

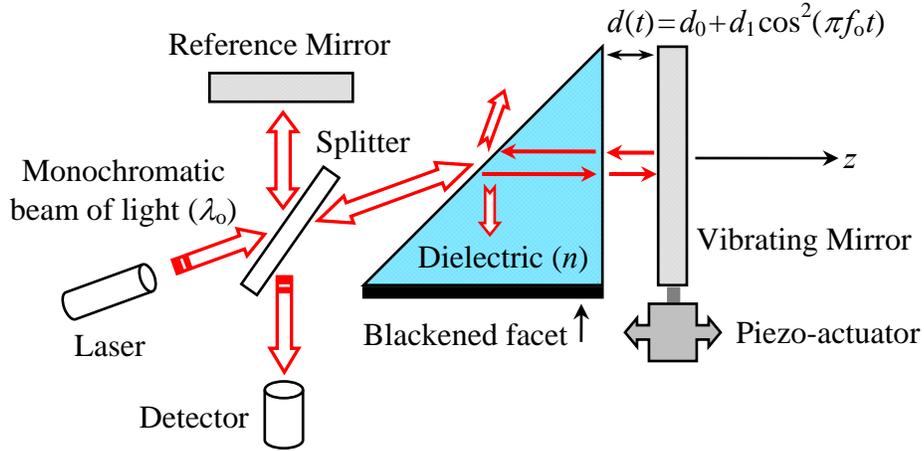

**Fig. 2** (Color online). A glass prism with blackened bottom is used to monitor the Doppler shift of a light beam reflected from a vibrating mirror behind a dielectric medium of refractive index $n$. A typical set of system parameters are $\lambda_o = 633$ nm, $f_o = 1.0$ kHz, $d_0 = d_1 = 1$ µm. The Michelson interferometer is set up to measure the phase $\phi(t)$ of the beam reflected from the vibrating mirror. The photodetector signal, $S(t) = S_0 + S_1 \cos\phi(t)$, is a periodic function of time $t$ (period $T = 1/f_o$). At each instant of time, the Doppler shift, given by $\Delta\omega(t) = \phi'(t)$, is proportional to the instantaneous velocity of the mirror, but the proportionality constant itself is a function of the gap-width $d(t)$ at that instant.

To measure the Doppler shift of the reflected beam from a moving mirror and to relate it to the instantaneous mirror velocity $V(t)$, to the corresponding gap-width $d(t)$, and to the refractive index $n$ of the dielectric slab, one could use the setup depicted in Fig. 2. Here the parallel-plate slab has been replaced by a 90° prism to avoid the use of antireflection coatings; note that the bottom of the prism is blackened to prevent multiple internal reflections. The CW light source used in the setup of Fig. 2 is quasi-monochromatic, with a coherence length greater than the optical path-length-difference between the two arms of the Michelson interferometer. We denote by $\phi(t)$ the phase difference between the two light beams arriving at the detector. Since the reference mirror is stationary, the time-dependence of the relative phase is entirely due to the vibrations of the mirror behind the glass prism. The photodetector signal, $S(t)=S_0+S_1\cos\phi(t)$, is a periodic function of time with a period $T=1/f_o$, where $f_o$ is the vibration frequency of the mirror. At each instant of time, the Doppler shift $\Delta\omega(t)$ is given by the time derivative $\phi'(t)$ of $\phi(t)$. As expected, this Doppler shift turns out to be proportional to the instantaneous velocity of the mirror, $V(t)=\pi f_o d_1 \sin(2\pi f_o t)$, although the proportionality constant also happens to vary with time due to its dependence on the gap-width $d(t)$; see Sec. 3 for a detailed explanation.

The system depicted in Fig. 2 thus enables the measurement of the Doppler shift of the light reflected from the mirror behind the glass prism as a function of the mirror velocity $V$, the gap-width $d$, and the refractive index $n$ of the dielectric material of the prism. Different mirrors having different phase-angles $\psi$ may also be used in the experiment, but since a change of the gap-width $d$ at the instant of measurement will have the same effect on the Doppler shift as a change of the phase-angle $\psi$, only one of these two variables would suffice.

**3. Doppler shift upon reflection from a moving mirror**. Our goal in the present section is to compute the Doppler shift of the reflected light in the system of Fig. 1 as a function of $\omega_o$, $n$, $d$, $\rho$ and $\psi$, when the mirror moves along the $z$-axis at a constant (and slow) velocity $V$. To this end, we fix the air-gap at $d$ and calculate the Fresnel reflection coefficient at the rear facet of the dielectric slab (i.e., at $z=0$) as a function of $d$. We find

$$r\exp(i\phi) = \frac{[(n-1)/(n+1)]+\rho\exp[i(\psi+4\pi d/\lambda_o)]}{1+[(n-1)/(n+1)]\rho\exp[i(\psi+4\pi d/\lambda_o)]}. \tag{1}$$

If the mirror is a 100% reflector, that is, if $\rho=1.0$, it is easy to show that the reflection coefficient $r$ of the glass slab at $z=0$ will also be equal to 1.0. In practice, $\rho$ will be close to but slightly less than unity, causing $r$ to vary ever so slightly with a change of the gap-width $d$ between the mirror and the slab. These variations, however, are going to be insignificant provided that $\rho$ is sufficiently large and the changes in $d$ are kept at a small fraction of one wavelength, $\lambda_o$.

The phase $\phi$ of the reflection coefficient at $z=0$ is readily obtained from Eq. (1), as follows:

$$\tan\phi = \frac{[2n/(n^2+1)]\sin(\psi+2\omega_o d/c)}{\{(n^2-1)(\rho^2+1)/[2(n^2+1)\rho]\}+\cos(\psi+2\omega_o d/c)}. \tag{2}$$

Let $d=d_o+Vt$, where $d_o$ is the width of the initial gap between the mirror and the dielectric slab, and $V$ is some small velocity with which the mirror moves away from the slab. The velocity $V$ is so small that during the time interval of interest, say, $0\leq t\leq t_o$, the gap $d$ will change only by a small fraction of the wavelength $\lambda_o$, thus keeping $r$ essentially constant and allowing one to approximate $\phi(t)$ as $\phi_0+\phi_1 t$, the first two terms in the Taylor series expansion of $\phi(t)$ around $t=0$. To simplify the notation, we define the new parameters $\alpha$, $\beta$, and $\psi_o$ as follows:

$$\alpha = 2n/(n^2+1), \tag{3a}$$

$$\beta = \frac{(n^2-1)(\rho^2+1)}{2(n^2+1)\rho}, \tag{3b}$$

$$\psi_o = \psi + 2\omega_o d_o/c = \psi + 4\pi d_o/\lambda_o. \tag{3c}$$

Noting that the argument of the sine and cosine functions in Eq.(2) may now be expressed as $\psi_o + 2(V/c)\omega_o t$, we will have

$$\phi_0 = \tan^{-1} \frac{\alpha \sin\psi_o}{\beta + \cos\psi_o}, \tag{4a}$$

$$\phi_1 = \frac{2(V/c)\omega_o \alpha (1 + \beta \cos\psi_o)}{(\beta + \cos\psi_o)^2 + \alpha^2 \sin^2\psi_o}. \tag{4b}$$

The time-dependence factor of the incident beam in Fig.1 is $\exp(-i\omega_o t)$, which must be multiplied by $\exp[i\phi(t)]$ to yield the corresponding factor for the reflected beam. The phase of the reflected beam thus becomes $\phi_0$, while its frequency changes to $\omega = \omega_o - \phi_1$. We thus have

$$\omega = \omega_o \left[ 1 - \frac{2(V/c)\alpha (1 + \beta \cos\psi_o)}{(\beta + \cos\psi_o)^2 + \alpha^2 \sin^2\psi_o} \right]. \tag{5}$$

This is the general formula for the Doppler shift of the reflected light associated with the slow, linear motion of the mirror away from the dielectric slab of Fig.1, at the instant when $d = d_o$. In the special case when $n = 1$ (i.e., no glass slab), we will have $\alpha = 1$, $\beta = 0$ and $\omega = \omega_o[1 - 2(V/c)]$, independent of the value of $\psi_o$, as expected for reflection from a mirror moving in free space at a non-relativistic speed.

When $\rho \approx 1$, we may write $\beta \approx (n^2 - 1)/(n^2 + 1)$ to a very good approximation, in which case Eq.(5), without further approximations, simplifies as follows:

$$\omega \approx \omega_o \left[ 1 - \frac{2nV/c}{1 + (n^2 - 1)\cos^2(\psi_o/2)} \right]. \tag{6}$$

Note that the Doppler shift of Eq.(6) is a strong function of the phase parameter $\psi_o$. Even when the mirror is initially in contact with the dielectric slab (i.e., when $d_o = 0$), the Doppler shift depends on the phase $\psi$ of the reflection coefficient of the mirror; only when $\psi = 180°$ do we find the familiar formula $\omega = \omega_o(1 - 2nV/c)$ with the mirror initially in contact with the slab. This is in stark contrast with the case of a submerged mirror within a dielectric liquid of refractive index $n$, where the standard Doppler formula, $\omega = \omega_o(1 - 2nV/c)$, would be independent of $\psi$.

**4. Radiation pressure on mirror initially at rest**. Suppose now that a single photon of energy $\hbar\omega_o$ arrives at the glass slab of Fig.1. Let the slab be extremely massive, so we will not have to consider any kinetic energy acquired by this slab as a result of its interactions with the photon. The mirror, however, while massive, is not excessively so, and we write its acquired momentum (upon reflection of the photon) as $p_{mirror} = mV$ and its corresponding kinetic energy as $\mathcal{E}_{mirror} = \tfrac{1}{2}mV^2$. Here $m$ is the mass and $V$ the final velocity of the mirror along the $z$-axis. Since the mirror is at rest before the arrival of the photon, and its velocity climbs to $V$ upon reflection, the Doppler shift associated with the reflected photon will correspond to the mirror's average velocity $V/2$. (We are making no approximations here; this is an exact statement.) Using conservation of energy and the Doppler formula of Eq.(6) (with $V$ replaced by $V/2$) we arrive at

$$\hbar\omega_o \approx \tfrac{1}{2}mV^2 + \hbar\omega_o\left[1 - \frac{nV/c}{1+(n^2-1)\cos^2(\psi_o/2)}\right]. \tag{7}$$

Eliminating $\hbar\omega_o$ from both sides of the above equation, then dividing by $V$, yields

$$p_{\text{mirror}} = mV \approx \frac{2n\hbar\omega_o/c}{1+(n^2-1)\cos^2(\psi_o/2)}. \tag{8}$$

This is the momentum given to the mirror upon reflection of a single photon through the dielectric slab. In the absence of the slab, of course, we will have $n=1$ and $p_{\text{mirror}} \approx 2\hbar\omega_o/c$, as expected. In the presence of the slab, the transferred momentum to the mirror depends on the refractive index of the glass as well as on the phase parameter $\psi_o$. If $\psi_o = 180°$, which is typical of metallic as well as most dielectric-stack mirrors, the mirror's momentum will become twice the Minkowski momentum of the photon, that is, $p_{\text{mirror}} \approx 2n\hbar\omega_o/c$. If, however, the phase parameter happens to be $\psi_o = 0°$, the mirror will acquire twice the Abraham momentum, $p_{\text{mirror}} \approx 2\hbar\omega_o/(nc)$. For intermediate values of the phase parameter, $p_{\text{mirror}}$ will cover the entire range of values between the Abraham and Minkowski limits.

**5. Direct calculation of the radiation pressure on the mirror**. We calculate the radiation pressure per photon on a 100% reflector having reflection coefficient $\exp(i\psi)$, placed a distance $d$ behind the glass slab of refractive index $n$, as shown in Fig.1. Let the incident light pulse be linearly polarized, having an $E$-field amplitude $E_o$ before entering the slab. Denoting by $Z_o = \sqrt{\mu_o/\varepsilon_o}$ the impedance of free space, where $\mu_o$ and $\varepsilon_o$ are the permeability and permittivity of free space, the total energy content of the pulse will be

$$\mathcal{E} = \frac{E_o^2 \tau}{2Z_o}. \tag{9}$$

Inside the slab, the $E$-field amplitude will be $E_o/\sqrt{n}$, while the $H$-field amplitude will be $\sqrt{n}E_o/Z_o$, as the anti-reflection coating ensures the continuity of the Poynting vector $\boldsymbol{S} = \boldsymbol{E}\times\boldsymbol{H}$ in passing from air into the dielectric slab. Denoting the $E$-field amplitude entering the gap at $z=0$ by $E_g$, the continuity of the tangential $E$-field at $z=0$ yields

$$E_g = \frac{[1+\exp(i\phi)]E_o/\sqrt{n}}{1+\exp(i\psi_o)} = \frac{2\sqrt{n}E_o}{(n+1)+(n-1)\exp(i\psi_o)}. \tag{10}$$

The momentum transferred to the mirror is thus given by

$$p_{\text{mirror}} = \varepsilon_o |E_g|^2 \tau = \frac{\varepsilon_o n E_o^2 \tau}{1+(n^2-1)\cos^2(\psi_o/2)}. \tag{11}$$

Considering that the pulse energy is given by Eq.(9), we may substitute for $E_o^2 \tau$ of a single photon the quantity $2Z_o \hbar \omega_o$, thereby transforming Eq.(11) into Eq.(8). (Here we have used the fact that $c = 1/\sqrt{\mu_o \varepsilon_o}$.) This confirms the validity of deriving the radiation pressure on the mirror from the Doppler shift of the reflected light.

Note that a mirror whose Fresnel reflection coefficient in vacuum is specified as $\rho \exp(i\psi)$, will exhibit a different reflection coefficient when submerged in a liquid of refractive index $n$. In fact, setting $d$ to zero in Eq.(1) yields an exact expression for the reflection coefficient $r \exp(i\phi)$ of the mirror as measured within the liquid. When the mirror happens to be 100% reflective, the magnitudes of both reflection coefficients (i.e., $\rho$ in vacuum and $r$ in the liquid) will be unity, while $\psi$ and $\phi$ will be related via Eq.(2) with $d$ set to zero, as follows:

$$\tan \phi = \frac{2n \sin \psi}{(n^2-1)+(n^2+1)\cos\psi}. \tag{12}$$

It is readily seen from the above equation that $\tan(\psi/2) = n \tan(\phi/2)$. Therefore, when $d=0$, Eq.(8) may be rewritten in terms of $\phi$, as follows:

$$p_{\text{mirror}} = 2[1+(n^2-1)\sin^2(\phi/2)]\hbar\omega_o/(nc). \tag{13}$$

The last equation is in agreement with the result obtained in [5], where the mirror was submerged in a liquid of refractive index $n$, and the phase of the Fresnel coefficient was specified as $\phi$ within the liquid (rather than as $\psi$, which is the corresponding phase in vacuum). The results of the preceding sections, therefore, are in complete agreement with our earlier results, which were obtained for a submerged mirror in a transparent liquid.

**6. Concluding remarks**. We have shown that the momentum transferred to a submerged mirror upon reflection of a single photon of frequency $\omega_o$ could be anywhere in the range from $2\hbar\omega_o/(nc)$ to $2n\hbar\omega_o/c$, depending on the phase angle $\psi$ of the mirror's reflection coefficient. We used a theoretical argument based on the Doppler shift of the reflected light to arrive at the above conclusion, even though the Doppler shift caused by radiation pressure is so tiny as to be practically immeasurable. Nevertheless, what is important is the proportionality coefficient between the Doppler shift and the mirror velocity, and this coefficient can be readily measured in a setup such as that shown in Fig.2. The proportionality coefficient was derived from theoretical considerations in Sec.3, where it was necessary to assume a small mirror velocity to ensure that multiple reflections within the gap would stabilize before the gap-width could change perceptibly. For high-$Q$ cavities and fast-moving mirrors, one expects the Doppler shift measured in the system of Fig.2 to depart from the predictions of Eq.(5). However, in the argument advanced in Sec.4 with regard to radiation pressure on a submerged mirror, the mirror velocity is exceedingly small and any deviation from Eq.(5) will be totally negligible.

Since the difference between the momenta of the incident photon (before entering the dielectric slab of Fig.1) and the reflected photon (upon emerging from the slab) is $2\hbar\omega_o/c$, it is clear that the difference between this value and $p_{\text{mirror}}$ of Eq.(8) must have been transferred to the glass slab. This can indeed be confirmed by a direct calculation of the Lorentz force of the light on the region of the slab where the incident and reflected beams overlap [16], and also on the antireflection coating at the entrance facet of the slab [11]. For the argument from Doppler shift presented in Sec.4 to be valid, it is important to recognize that, while the slab picks up a net momentum of $2(\hbar\omega_\text{o}/c) - p_\text{mirror}$ from each photon, its acquired kinetic energy will be negligible, simply because its mass may be assumed to be much greater than that of the mirror. This, in fact, was our requirement for the validity of the energy balance equation, Eq. (7).

In a recent paper [17], Kemp questioned the validity of Eq. (13), which we had derived earlier from consideration of the Lorentz force on submerged mirrors [5]. Kemp's argument is based on an analysis of the Doppler shift of the reflected light, but his energy balance equation is incomplete in that the energy required to drag the submerging liquid along with the mirror has been ignored. In Kemp's analysis, unlike ours in Sec. 4, the mirror remains in contact with the surrounding liquid while being driven forward by radiation pressure. The mirror must therefore drag the liquid molecules along as it moves forward. Had Kemp taken into account the energy needed to bring these molecules along – molecules upon which acts the Lorentz force within the fringes produced by interference between incident and reflected beams – his analysis would have yielded the same mirror momentum as that given by Eq. (13).

In another recent paper [18], Barnett invoked a thought experiment (originally suggested by Milonni *et al* [19,20]) to argue in favor of the Minkowski momentum for a photon inside a transparent medium. The thought experiment involves the absorption of a photon by an atom submerged in a liquid dielectric. The Doppler shift of the photon (as seen by the moving atom) in conjunction with energy conservation is used to arrive at the Minkowski momentum for the photon. As was the case with Kemp's analysis, we believe the energy balance equation used by Milonni *et al* is incomplete. Had these authors taken into consideration the energy needed to drag along the liquid molecules – which molecules are subject to the Lorentz force at the trailing edge of the wavepacket – they would have found a momentum equal to $\tfrac{1}{2}(n+n^{-1})\hbar\omega_\text{o}/c$ for individual photons inside a medium of refractive index *n*; see [21] for a detailed discussion of this point.

The bottom line is that there is perfect agreement between experiments that indicate a recoil momentum of $n\hbar\omega_\text{o}/c$ inside dielectric media [1-3,22-25], and theoretical arguments that show a photon momentum of $\tfrac{1}{2}(n+n^{-1})\hbar\omega_\text{o}/c$ that is divided between an electromagnetic part ($\hbar\omega_\text{o}/nc$) and a mechanical part [$\tfrac{1}{2}(n-n^{-1})\hbar\omega_\text{o}/c$]. It is immaterial whether theory relies on direct calculations of the Lorentz force, as in [26], or is based on an analysis of the Doppler shift, as in the present paper. The computed recoil momentum in every case agrees with experimental observations without violating the conservation laws, as the balance of energy and momentum is picked up by the host dielectric.

**Acknowledgement**. The author is grateful to Brian Anderson, Stephen Barnett, Poul Jessen, Khanh Kieu, Henri Lezec, Ewan Wright, and Armis Zakharian for helpful discussions.

**References**

1. R. V. Jones and J.C.S. Richards, *Proc. Roy. Soc. A* **221**, 480 (1954).
2. R. V. Jones and B. Leslie, "The measurement of optical radiation pressure in dispersive media," Proc. Roy. Soc. London, Series A, **360**, 347-63 (1978).
3. R. V. Jones, "Radiation pressure of light in a dispersive medium," Proc. Roy. Soc. London Ser. A. **360**, 365-71 (1977).
4. F. N. H. Robinson, "Electromagnetic stress and momentum in matter," Phys. Rep. **16**, 313-54 (1975).
5. M. Mansuripur, "Radiation pressure on submerged mirrors: Implications for the momentum of light in dielectric media," Optics Express **15**, 2677-2682 (2007).
6. L. Landau, E. Lifshitz, *Electrodynamics of Continuous Media*, Pergamon, New York, 1960.
7. J. D. Jackson, *Classical Electrodynamics*, 3rd edition, Wiley, New York, 1998.
8. J. P. Gordon, "Radiation forces and momenta in dielectric media," Phys. Rev. A **8**, 14-21 (1973).
9. R. Loudon, "Theory of the radiation pressure on dielectric surfaces," J. Mod. Opt. **49**, 821-38 (2002).

10. R. Loudon, "Radiation pressure and momentum in dielectrics," Fortschr. Phys. **52**, 1134-40 (2004).
11. M. Mansuripur, "Radiation pressure and the linear momentum of the electromagnetic field," *Optics Express* **12**, 5375-5401 (2004).
12. S. M. Barnett and R. Loudon, "On the electromagnetic force on a dielectric medium," *J. Phys. B: At. Mol. Opt. Phys.* **39**, S671-S684 (2006).
13. S. M. Barnett and R. Loudon, "The enigma of optical momentum in a medium," Philos. Trans. R. Soc. London Ser. A **368**, 927-39 (2010).
14. C. Baxter and R. Loudon, "Radiation pressure and the photon momentum in dielectrics," J. Mod. Opt. **57**, 830-42 (2010).
15. P. W. Milonni and R. W. Boyd, "Momentum of light in a dielectric medium," Advances in Optics and Photonics **2**, 519-53 (2010).
16. M. Mansuripur, "Momentum of the electromagnetic field in transparent dielectric media," Proc. SPIE **6644**, 664413, pp 1-10 (2007).
17. B. A. Kemp, "Resolution of the Abraham-Minkowski debate: Implications for the electromagnetic wave theory of light in matter," J. Appl. Phys. **109**, 111101 (2011).
18. S. M. Barnett, "Resolution of the Abraham-Minkowski dilemma," Phys. Rev. Lett. **104**, 070401 (2010).
19. P. W. Milonni and R. W. Boyd, "Recoil and photon momentum in a dielectric," Laser Phys. **15**, 1432-38 (2005).
20. D. H. Bradshaw, Z. Shi, R. W. Boyd, and P. W. Milonni, "Electromagnetic momenta and forces in dispersive dielectric media," Opt. Commun. **283**, 650-656 (2010).
21. M. Mansuripur, "Solar Sails, Optical Tweezers, and Other Light-Driven Machines," in *Tribute to Joseph W. Goodman*, edited by H. J. Caulfield and H. H. Arsenault, Proc. SPIE **8122**, 81220D, pp 1-13 (2011).
22. A. Ashkin and J. Dziedzic, "Radiation pressure on a free liquid surface," Phys. Rev. Lett. **30**, 139-142 (1973).
23. I. Brevik, "Experiments in phenomenological electrodynamics and the electromagnetic energy-momentum tensor," Phys. Rep. **52**, 133-201 (1979).
24. G. K. Campbell, A. E. Leanhardt, J. Mun, M. Boyd, E. W. Streed, W. Ketterle, and D. E. Pritchard, "Photon recoil momentum in dispersive media," Phys. Rev. Lett. **94**, 170403 (2005).
25. R. N. C. Pfeifer, T. A. Nieminen, N. R. Heckenberg, and H. Rubinsztein-Dunlop, "Colloquium: momentum of an electromagnetic wave in dielectric media," Rev. Mod. Phys. **79**, 1197-1216 (2007).
26. M. Mansuripur and A. R. Zakharian, "Whence the Minkowski momentum?" Opt. Commun. **283**, 3557-63 (2010).